\documentclass[symmetry,article,submit,pdftex,moreauthors]{Definitions/mdpi}
\preto{\abstractkeywords}{\nolinenumbers}

\firstpage{1}
\pubvolume{1}
\issuenum{1}
\articlenumber{0}
\pubyear{2025}
\copyrightyear{2025}
\datereceived{ }
\dateaccepted{ }
\datepublished{ }
\hreflink{https://doi.org/} 

\newcommand{\eref}[1]{Equation~(\ref{#1})}
\newcommand{\fref}[1]{Figure~\ref{#1}}
\newcommand{\aref}[1]{Appendix~\ref{#1}}

\Title{Zak Phase Dislocations in Trimer Lattices}

\TitleCitation{Zak Phase Dislocations in Trimer Lattices}

\Author{Tileubek Uakhitov$^{1}$, Abdybek Urmanov$^{1}$, Serik E. Kumekov$^{1,2}$ and Anton S. Desyatnikov$^{1,*}\orcidA{}$}

\AuthorNames{Uakhitov T., Urmanov A., Kumekov S.E. and Desyatnikov A.S.}

\isAPAStyle{%
       \AuthorCitation{Lastname, F., Lastname, F., \& Lastname, F.}
         }{%
        \isChicagoStyle{%
        \AuthorCitation{Lastname, Firstname, Firstname Lastname, and Firstname Lastname.}
        }{
        \AuthorCitation{Uakhitov T.; Urmanov A.; Kumekov S.E.; Desyatnikov A.S.}
        }
}

\address{%
$^{1}$ \quad Department of Physics, School of Sciences and Humanities, Nazarbayev University, 53 Kabanbay Batyr Ave., Astana 01000, Kazakhstan\\
$^{2}$ \quad Satbayev Kazakh National Technical University, 22a Satbayev Str., Almaty 050000, Kazakhstan}

\corres{Correspondence: anton.desyatnikov@nu.edu.kz}

\abstract{Wave propagation in periodic media is governed by energy-momentum relation and geometric phases characterizing band topology, such as Zak phase in one-dimensional lattices. We demonstrate that in the off-diagonal trimer lattices Zak phase carries quantized screw-type dislocations winding around degeneracies in parameter space. If the lattice evolves in time periodically, as in adiabatic Thouless pump, corresponding closed trajectory in parameter space is characterized by a Chern number equal the negative total winding number of Zak phase dislocations enclosed by the trajectory. We discuss correspondence between bulk Chern numbers and the edge-states in a finite system evolving along various pumping cycles.}


\begin{document}

\section{Introduction}

Geometric Berry phase~\cite{BerryPhase} defines topological properties of wave transport in periodic media in condensed matter~\cite{RMPBerry,RMPHasanKane,RMPtopins} and photonics~\cite{TP14,TP19,TP20}. In one-dimensional lattices the Berry phase, introduced by Zak~\cite{Zak}, is associated with adiabatic evolution in the Brillouine zone. Zak phase $Z$ was experimentally measured with cold atoms~\cite{ZakBloch} in optical dimer lattices, realizing Su-Schrieffer-Heeger (SSH)~\cite{SSH} and Rice-Mele~\cite{RM} models. The SSH system exhibit topological transition from trivial ($Z=0$) to nontrivial ($Z=\pi$)~\cite{ZakSSH}, the latter bulk property is manifested in a finite open system by the appearance of a pair of edge-states inside the spectral gap. Other lattices without inversion-symmetry may posses nonzero phase $Z$ not quantized by $\pi$~\cite{Zak}; the role of this geometric phase remains obscure.

Here we demonstrate the appearance of screw-type Zak phase dislocations in parameter domain of the off-diagonal trimer lattices, i.e. one-dimensional periodic lattices with three equivalent atoms in a unit cell. The family of trimer lattices is spanned by three hopping parameters: intra-cell $J_{1,2}$ and inter-cell $J_3$. The lattice is inversion-symmetric for $J_1=J_2$ with vanishing Zak phase $Z=0$, which may suggest that the system is topologically trivial. However, several studies~\cite{diagonal12,ChaoYuri,trimerized17,ScRep17_trimer,Brazil} have shown the presence of paired and unpaired chiral edge-states and piecewise-continuous Zak phase~\cite{ScRep17_trimer} for $J_1\ne J_2$. Our analysis reveals the winding of Zak phase around the degeneracy in parameter space, $J_1=J_2=J_3$, where the gaps close in monatomic lattice.

Phase dislocations were introduced by Nye and Berry~\cite{BerryDislocations} as generic feature of waves of any nature, e.g. in optics they are associated with quantized vortices~\cite{Soskin,DesyatnikovKivsharTorner,DennisOHolleran} and orbital angular momentum of light~\cite{OAM}. As pointed by Berry~\cite{BerryPhase}, the degeneracies act as organising centres for phase changes, and we demonstrate below how Zak phase dislocations define the properties of trimer lattices.

Geometric phase manifests itself in the evolution of a system in parameter space~\cite{BerryPhase}. If the evolution is periodic, the adiabatic driving leads to quantization of the current, predicted by Thouless~\cite{Thouless}. The adiabatic Thouless pumps were demonstrated with cold atoms in optical lattices slowly changing in time~\cite{pumpSSHcold16,pumpSSHcold16_2} and with optical waveguides slowly changing in the direction of propagation~\cite{diagExperPRL12}. The latter system offers novel possibilities in designing periodic media~\cite{Longhi,GaranovichLonghiKivshar}, from Floquet topological insulators~\cite{NATMoti} to nonlinear~\cite{nonlinearTP} and active~\cite{activeNano} topological photonic structures~\cite{TP20}. Adiabatic cyclic parameter can be seen as an additional ``synthetic'' degree of freedom, thus effectively increasing the dimension of the system~\cite{Nat18a,Nat18b}. The ideas of the synthetic dimensions in photonics~\cite{synthetic} now extend to the modal subspaces~\cite{NATMotiSynth,ScienceFan20} and emulation of up to seven-dimensional hyper-lattices~\cite{Andreys}.

One-dimensional photonic lattices modulated in synthetic dimension relate to Aubry-Andr\'{e}-Harper (AAH) model~\cite{AAH,Harper,diagExperPRL12}, mapped to a two-dimensional integer quantum Hall system~\cite{RMPHasanKane,RMPtopins}. The latter is characterized by an integer topological index, the Chern number, determined by the Berry curvature of Bloch bands~\cite{RMPBerry}. In our trimerized commensurate off-diagonal AAH model the Chern numbers for the three bands are $\{-1,2,-1\}$ ~\cite{ChaoYuri,Brazil}. Most interesting is the topological transition to double negative Chern numbers, $\{2,-4,2\}$, predicted numerically for the modulation depth exceeding a threshold value~\cite{ChaoYuri}.

We demonstrate below that the band Chern number in modulated trimer lattices can be explicitly calculated as the negative total winding number of Zak phase dislocations enclosed by a loop in parameter space. We explain the topological transitions as the crossing of dislocations by adiabatic loop, changing the winding numbers. We suggest a general scheme to construct adiabatic protocols, also with large Chern numbers up to $\{-4,8,-4\}$. We conclude with a version of the bulk-boundary correspondence, linking Chern numbers with the sequence of edge-states accompanying evolution.

The paper is organised as follows. In the next Section~\ref{S2} we introduce our model, describe relevant experimental settings, and analyse the parameter domain of the off-diagonal trimer lattices. We demonstrate the winding of geometric Zak phase in the parameter space of the infinite periodic lattices and its relation to the appearance of paired and unpaired edge-states in the spectral gaps of finite lattices. Section~\ref{S3} is devoted to the study of adiabatic Thouless pumps in the slowly varying lattices. We establish analytically the relation between Zak phase winding numbers, Chern numbers of adiabatic loops, and corresponding sequences of the edge-states. Section~\ref{S4} concludes the paper and the Appendices clarify technical aspects of our derivations.

\section{Trimer lattice and Zak phase}\label{S2}

\begin{figure}[h]
\centering
\includegraphics[width=0.7\columnwidth]{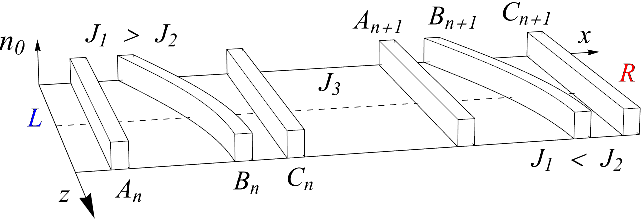}%
\caption{\label{Fig0} Illustration of a trimer lattice in photonics~\cite{Alex}. Plotted is the refractive index $n_0(x,z)$ for two unit cells, numbered $n$ and $n+1$, of a trimer lattice with the intra-cell coupling strengths $J_{1,2}(z)$ slowly changing with propagation distance $z$. Dashed line corresponds to the inversion-symmetric lattice with $J_1=J_2$. We count the unit cells from the left (L) to the right (R) edges of the lattice.}
\end{figure}

We introduce trimer lattices using their potential experimental implementation in photonics as a basic example~\cite{Alex}; \fref{Fig0} illustrates the concept. Namely, we consider identical single-mode optical waveguides arranged periodically close enough for the tails of their transverse modes to overlap. The overlap allows for the energy exchange, thus coupling the nearest neighbours. The overlap decays exponentially with the distance between waveguides; varying the separation between waveguides allows to control the three coupling parameters $J_m$, $m=1,2,3$. In particular, the adiabatic modulation of these couplings with the propagation distance $z$ is analogous to the tight-binding model with hopping amplitudes slowly varying in time.

Complex amplitudes of the modes in each waveguide are labeled with the three sublattice names and the unit cell number $n$, see~\fref{Fig0},
\begin{eqnarray}\label{ABC}
&&i \tfrac{d}{dz}A_n + J_1 B_n +J_3 C_{n-1}=0,\nonumber\\
&&i \tfrac{d}{dz}B_n + J_1 A_n +J_2 C_n=0,\\
&&i \tfrac{d}{dz}C_n + J_2 B_n +J_3 A_{n+1}=0.\nonumber
\end{eqnarray}
We distinguish two systems: (i) an infinite periodic lattice with $n\in(-\infty,\infty)$, and (ii) finite lattice of $N$ unit cells, $n\in (1,N)$, and open boundary conditions, $A_n=B_n=C_n=0$ for $n<1$ and for $n>N$. Note the scaling invariance of~\eref{ABC}: introducing an arbitrary propagation scale $z_0$, i.e. $z\to z/z_0$, is equivalent to the scaling $J_m\to z_0 J_m$. In other words, the longer propagation in the waveguides is similar to the increased strength of the couplings between them.

Our basic model is an infinite periodic lattice with constant hopping parameters. Applying the Fourier transform, $\{A_n(z),B_n(z),C_n(z)\}=\int^\pi_{-\pi} dq \{a(q),b(q),c(q)\}\exp(iqn-i\omega(q) z)$, we describe the lattice with the Bloch states, $\psi=\{a,b,c\}^T$, and the crystal quasi-momentum $q=[-\pi,\pi]$, defined in a periodic Brillouin zone, $\psi(q+2\pi)=\psi(q)$. These states satisfy the eigenvalue problem, $\omega\psi=H\psi$, with the Bloch Hamiltonian
\begin{equation}
H=-
\left(
  \begin{array}{ccc}
    0 & J_1 & J_3 e^{-iq}\\
    J_1 & 0 & J_2 \\
    J_3e^{iq} &J_2 & 0 \\
  \end{array}
\right).
\label{Hq}
\end{equation}
The dispersion relation $\omega(q)$ is derived as the solution to the cubic equation,
\begin{equation}
\omega (\omega^2-J^2)=-2p\cos q,
\label{disp}
\end{equation}
here $J^2=J_1^2+J_2^2+J_3^2$ and $p=J_1J_2J_3$; we will use the band index $\mu=1,2,3$ for the three roots $\omega_1\le\omega_2\le\omega_3$. Corresponding normalized Bloch wavefunctions
\begin{equation}
\psi_\mu(q) =\Psi_\mu\left(
\begin{array}{c}
-\omega_\mu J_1+J_2J_3 e^{-iq} \\
\omega_\mu^2-J_3^2 \\
-\omega_\mu J_2+J_1J_3 e^{iq}
\end{array}\right),
\label{psi}
\end{equation}
here $\Psi_\mu^{-2}=(3\omega_\mu^2-J^2)(\omega_\mu^2-J_3^2)$.

\begin{figure}[t]
\centering
\includegraphics[width=0.95\columnwidth]{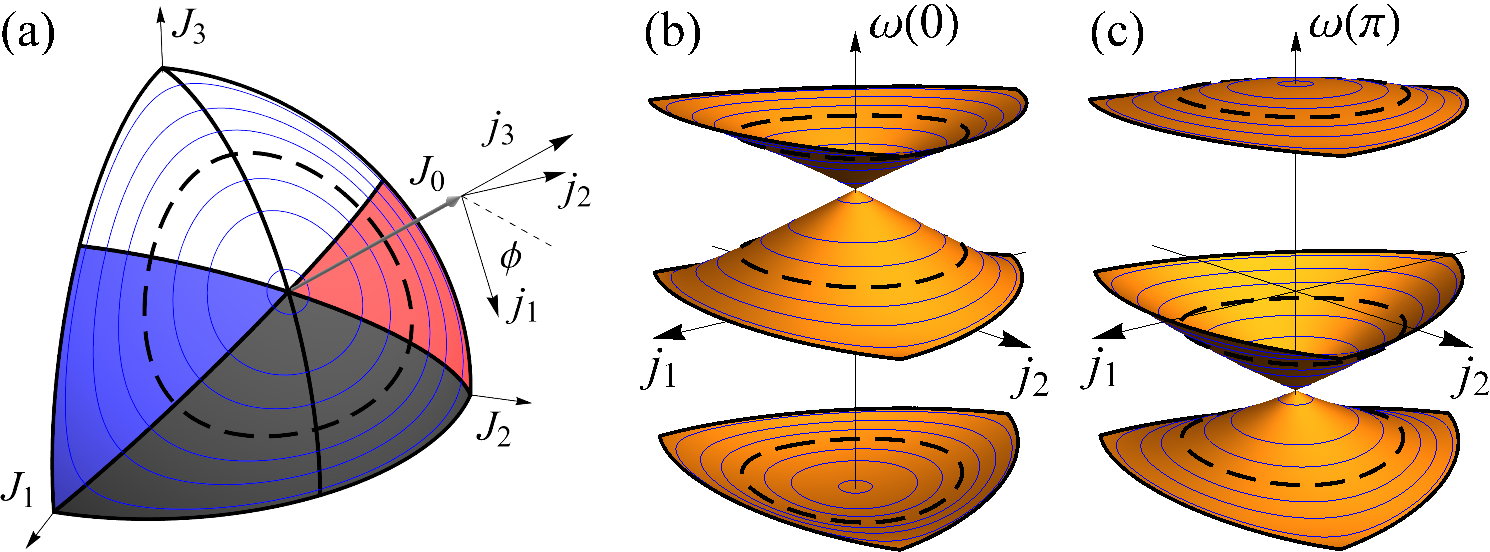}%
\caption{\label{Fig1} ({\bf a}) Parameter domain $J_{1,2,3}\ge0$ on the sphere $J$. Different sectors shaded in accordance with the number of edge states in a finite lattice: the gray-shaded sector has no edge-states, the blue and white sectors contain left-edge states, and the right-edge states appear in red and white sectors, thus the white sector contains both edge-states. The dispersion~\eref{disp} of an infinite lattice is plotted for ({\bf b}) $q=0$ and ({\bf c}) $q=\pi$, the gaps close at the degeneracy axis $J_0\equiv j_3$ with corresponding bands forming Dirac cones. The contour lines correspond to parameter $p=0.01, 0.04, 0.07, 0.1$ ({\it dashed}), $0.13, 0.16$, and $0.19$ (closest to the maximal value $p_{\max} =3^{-3/2}$ at the degeneracy $J_0$).}
\end{figure}

Parameter $J$ can be scaled to $1$ by choosing $z_0=J$ in~\eref{ABC} and transforming $J_m\to JJ_m$ and $\omega\to J\omega$, thus reducing the dimensions of parameter space from 3 to 2. On the sphere $J=const$ the only parameter which defines the energy $\omega(q)$ in~\eref{disp} is $p=J_1 J_2 J_3$. Therefore, the dispersion relation $\omega (q)$ does not change if we follow a line $p=const$ on the sphere $J$. These ``dispersionless'' trajectories are closed loops around the degeneracy axis $J_0 = J\left\{1,1,1\right\}/\sqrt{3}$ (monatomic lattice); several such level contour lines of the surface $p=const$ are drawn on a sphere in~\fref{Fig1}a. The loops originate from a degeneracy point $J_0$ with a maximal value $p_{\max} =3^{-3/2}$ and, as $p\to 0$, they converge to the boundary of the spherical triangle in~\fref{Fig1}a where one or two hopping parameters vanish, at the edges or vertices, respectively.

We would like to separate two degrees of freedom: the spectral parameter $p$ and the orthogonal dimension along the contours in~\fref{Fig1}. Evidently, while the second cyclic parameter does not influence $\omega(q)$, it modifies the eigenvectors~\eref{psi} and generates a family of states. In order to emphasise the role of the cyclic parameter, which does not influence energy-momentum relation, we introduce the new parameters $\{j_1,j_2,j_3\}$ by identifying $j_3$ with the degeneracy axis $J_0$; we identify $j_1$ ($j_2=0$) with the inversion-symmetric lattice $J_1=J_2$; and we choose azimuthal angle $\phi =\tan^{-1} (j_2/j_1)$ as the cyclic parameter, see~\aref{Ap1}.

The new ``coordinates'' $j_m$ are also useful in visualisation of the dispersion relation~\eref{disp} in~\fref{Fig1}b,c. The degeneracy axis $J_0\equiv j_3$ is placed at the center of the spherical triangle, $j_1=j_2=0$, where the three bands join into a single band of a monatomic lattice. This is not an energy-momentum relation, of course, because the eigenvalue $\omega$ is parameterized here by the coupling parameters. Therefore, the Dirac cones in~\fref{Fig1}b,c do not relate directly to the Dirac cones in, e.g., graphene or other two-dimensional lattices~\cite{LeykamDirac_2}. However, their appearance is not incidental and demonstrates the relation of the families of one-dimensional lattices to their two-dimensional counterparts. We will demonstrate that the additional parameter of a one-dimensional lattice family can be treated as a synthetic dimension along which a quasi-momentum can be introduced thus expanding the family into a effectively two-dimensional Floquet lattice.

In the following we study the family of trimer lattices in the parameter domain $J_m\ge0$ with the focus on the parameterisation by the cyclic parameter $\phi$, tracing a particular contour along $p=0.1$ dashed curves in~\fref{Fig1} as a representative example. Similar to~\fref{Fig1} we will compare the appearance of the edge-states in a finite lattice with the bulk property of the eigenstates in the bands of an infinite lattice, namely the Zak phase.

Zak phase is defined for each band as~\cite{Zak}
\begin{equation}\label{Zak}
Z_\mu = \int_{-\pi}^{\pi} A_{\mu q} dq,
\end{equation}
with Berry connection~\cite{BerryPhase,RMPBerry}
\begin{equation}\label{Aq}
A_{\mu q}=i\langle\psi_\mu|\partial_q \psi_\mu\rangle = (J_2^2-J_1^2)J_3^2\Psi_\mu^2,
\end{equation}
here we used the explicit solution for the Bloch spinor in~\eref{psi}. $A_{\mu q}$ manifestly vanishes in the inversion-symmetric lattice $J_1=J_2$, marking it as topologically trivial~\cite{Zak}, see~\aref{Ap21}. However, our further analysis challenges this conclusion.

\begin{figure}[t]
\includegraphics[width=1\columnwidth]{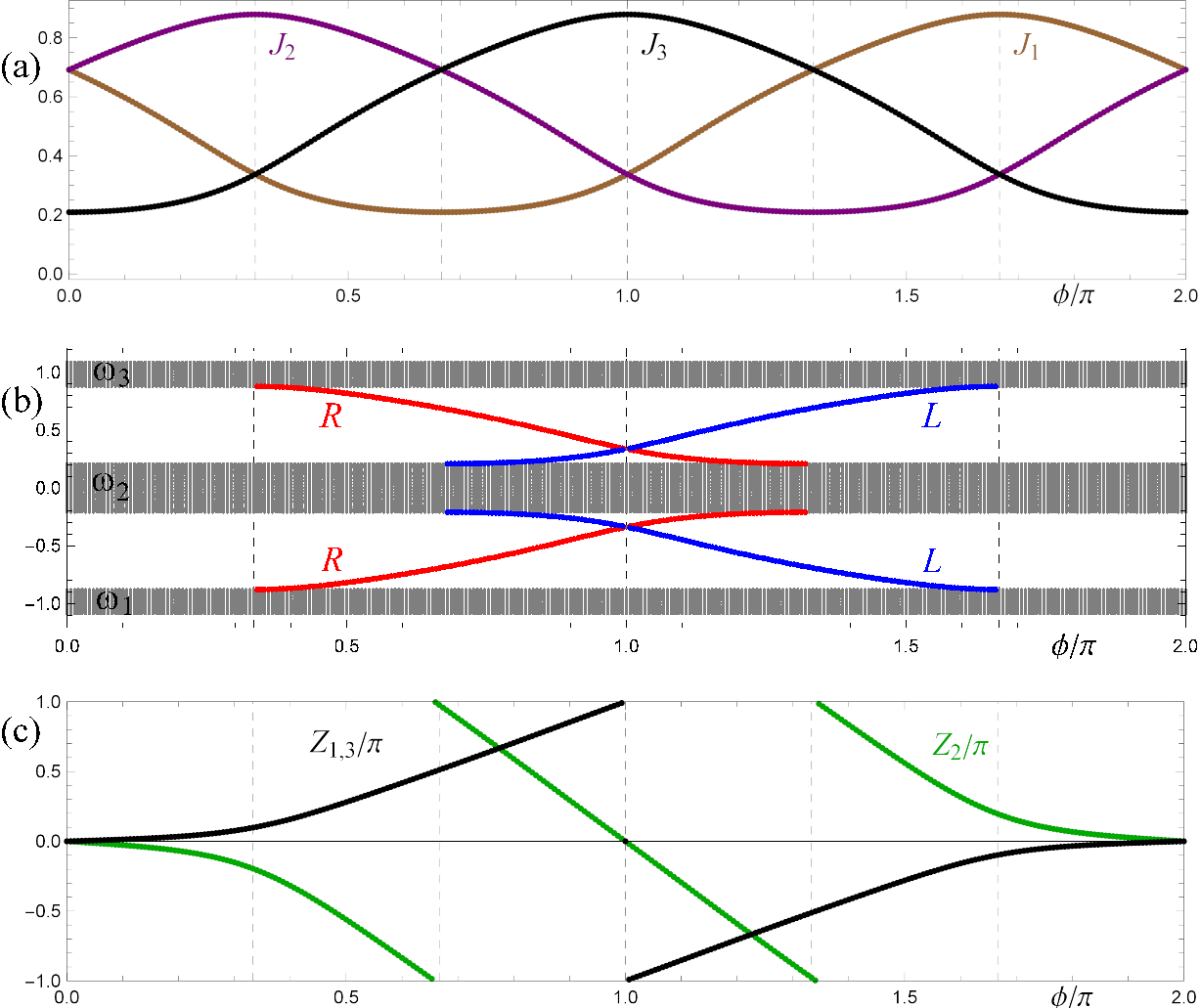}%
\caption{\label{Fig2} Family of states along the dashed curve $p=0.1$ in~\fref{Fig1}. ({\bf a}) Hopping parameters $J_m(\phi)$ help an eye to connect the dashed grid lines at $\phi= \pi s/3$ ($s=1,2,4,5$) with noncentered inversion symmetries $J_3=J_1$ and $J_3=J_2$~\cite{inversionPRB19}. ({\bf b}) Numerically obtained spectra for a chain of $N=50$ unit cells with open boundaries. The bands are shaded gray, the gap states are marked (R, red) for the right-edge and (L, blue) for the left-edge. ({\bf c}) Zak phases~\eref{Zak} are plotted mod$ \left[-\pi,\pi \right]$; note that $Z_\mu\equiv0$ for $J_1=J_2$ at $\phi=0,\pi$.}
\end{figure}

In~\fref{Fig2} we study the finite open system solved numerically by diagonalisation of the discretised Hamiltonian in~\eref{Hq}: as expected, the bands are completely flat in synthetic dimension $\phi$, i.e. for every band state $\partial_\phi \omega_\mu(q)=0$. Flat spectral bands appear naturally in systems with particle-hole symmetry with odd number of bands~\cite{LeykamDirac_2,LeykamDirac}, and they exhibit interesting sensitivity to disorder and interactions~\cite{flach,flach_2}. Suppression of dispersion in synthetic dimension may enhance Thouless pumping~\cite{Chao19}; here it is achieved simply by choosing particular modulation of hopping parameters in~\fref{Fig2}a.

The most revealing topological property is the appearance of edge states in the spectral gaps of insulators~\cite{RMPHasanKane,RMPtopins}. 
Edge modes in~\fref{Fig2}b bifurcate from bands at specific values of $\phi$, connected with noncentered inversion symmetries of the lattice~\cite{inversionPRB19,SSH3BulkEdgeCorr,SSH3extended,SSH3IRM,Symmetry24}. The domains of their existence~\cite{Brazil} can be summarized as follows: (R) right-edge modes for $J_3>J_1$ and (L) left-edge modes for $J_3>J_2$. This definition is sufficient to describe the whole parameter domain $J_m>0$ in~\fref{Fig1}a. Indeed, an overlap of both domains, $J_3>J_{1,2}$, contains both edge modes and it is marked white in~\fref{Fig1}a. If none of the conditions is satisfied, $J_3<J_{1,2}$, there are no edge-states and corresponding area in~\fref{Fig1}a is shaded gray. The red-colored sector $\pi/3<\phi<2\pi/3$ ($J_1<J_3<J_2$) corresponds to a single right-edge state in each spectral gap while the sector with a single left-edge state, $4\pi/3<\phi<5\pi/3$ ($J_2<J_3<J_1$), is shaded blue.

At the same time, the inversion-symmetric lattice exhibit sharp transition along $J_1=J_2$ from a domain with no edge states at $J_3<J_1=J_2$ ($\phi=0$) to a symmetric pair of gap states on both edges at $J_3>J_1=J_2$ ($\phi=\pi$). This transformation resembles closely the topological transition in SSH model~\cite{SSH,ZakSSH}, accompanied by a change of Zak phase from $0$ to $\pi$. However, as we demonstrated above, in our model the Zak phase is trivial in the whole domain with inversion symmetry, $J_1=J_2$, which naturally includes topologically trivial monoatomic lattice with $J_1=J_2=J_3$. Nevertheless, we observe in~\fref{Fig2}c that any other crossing of degeneracy from $\phi$ to $\phi+\pi$ is accompanied by Zak phase jumps close to $\pi$ for $Z_{1,3}$ and $2\pi$ for $Z_2$, thus confirming observations of Ref.~\cite{ScRep17_trimer}.

\begin{figure}[h]
\centering
\includegraphics[width=0.95\columnwidth]{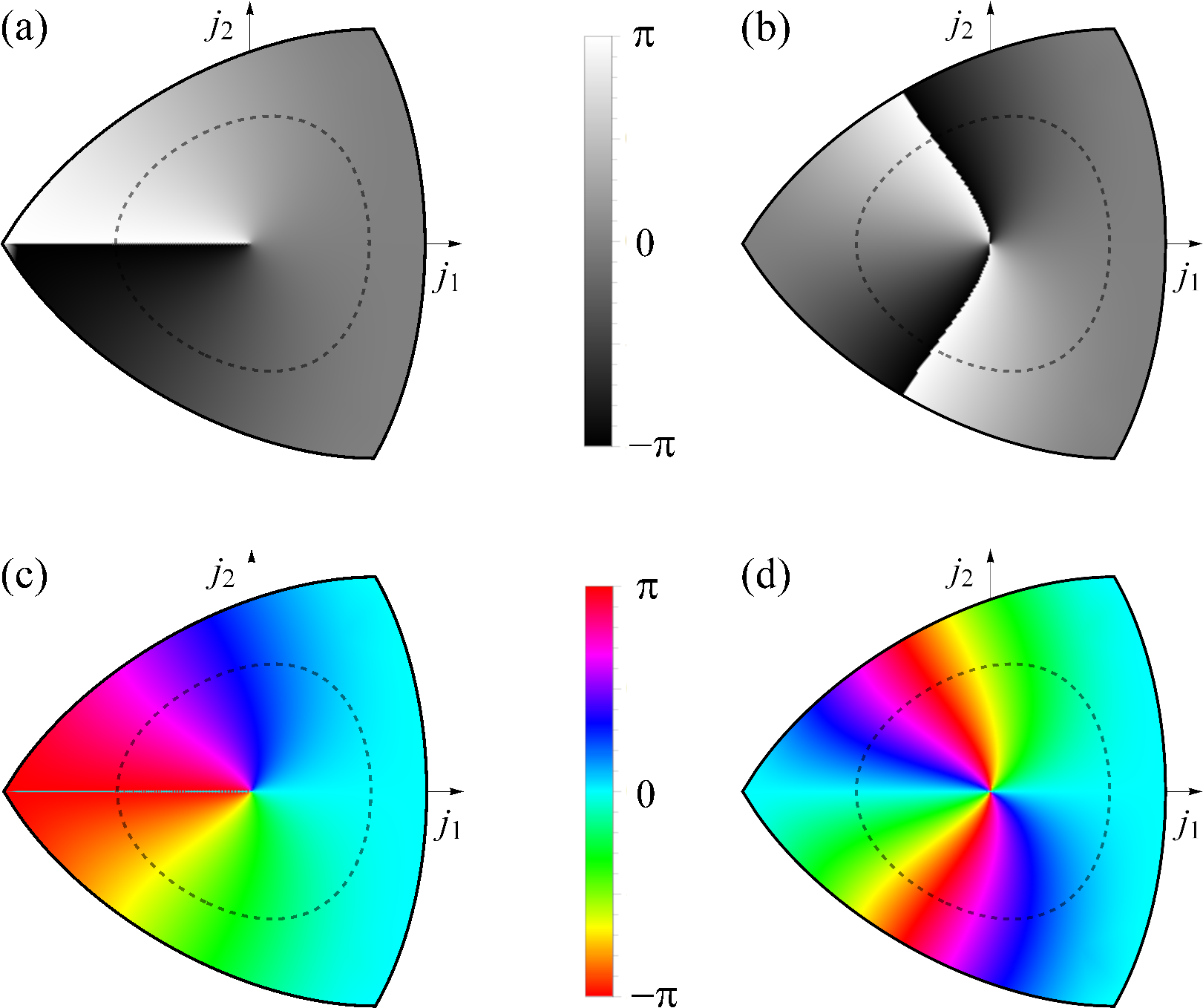}%
\caption{\label{Fig3} Zak phases~\eref{Zak}: ({\bf a,c}) $Z_{1,3}$ with winding number $l_{1,3}=+1$ and ({\bf b,d}) $Z_2$ with winding number $l_2=-2$. The dashed contour $p=0.1$ corresponds to~\fref{Fig2}c. The branch cuts are evident in gray scale in ({\bf a,b}) and they almost vanish if we plot $Z_{1,3}$ mod$[0,2\pi]$ and $Z_2$ mod$[-2\pi,2\pi]$, or use cyclic colormap, such as hue in ({\bf c,d}). The data plotted in ({\bf a,b}) and ({\bf c,d}) are exactly the same, only the colormaps are different.}
\end{figure}

The discontinuities of Zak phases in~\fref{Fig2}c and Ref.~\cite{ScRep17_trimer} is a feature familiar as branch cuts in twisted phases around screw-type wave dislocations~\cite{BerryDislocations}; in complex fields phase $\sim e^{il\phi}$ is quantized by an integer winding number $l$, often called topological charge. In other words, while the straight path from $\phi$ to $\phi+\pi$ in~\fref{Fig1}b,c involves crossing degeneracy with gaps closure, the same phase jump is gained on the path around the degeneracy point. Therefore, we calculate Zak phases in the whole parameter domain; ~\fref{Fig3} reveals their twists around the degeneracy with well defined winding numbers, $l_\mu=\{1,-2,1\}$. In contrast to complex fields, e.g. optical vortices~\cite{Soskin}, there is no singularity at the origin $j_{1,2}=0$ because Zak phases are well defined zeros along the $j_1$ axis in~\fref{Fig3}.

It is interesting to connect the winding numbers $l_\mu$ with the Dirac cones in~\fref{Fig1}b,c. While the first and the third bands contain only one Dirac cone, at $q=\pi$ and $q=0$, respectively, the second band is connected to both Dirac cones with opposite circulation~\cite{BerryPhase}, thus $l_2=-2$. We recall the ``wormhole'' argument by Haldane~\cite{Haldane04}, namely that the two Dirac singularities couple the bands by a ``Berry flux loop'' were Berry curvature flux passes from one band to the other through the first Dirac point, then returns through the second.

\section{Thouless pump and Chern numbers}\label{S3}

So far, we considered a family of lattices in the parameter space. A natural question is the role of collective features, such as Zak phase dislocations, in the adiabatic evolution along the family sequence~\cite{ChaoYuri,ScRep17_trimer,Brazil,diagExperPRL12}. We assume an arbitrary closed trajectory $\vec{\tau}$, parameterized by its length $\tau=|\vec{\tau}|$ and given by $J_m(\tau)$ functions, governing the energy~\eref{disp} and eigenstates~\eref{psi} along the loop. The Chern number is defined as the surface integral over two-dimensional Berry curvature~\cite{BerryPhase,RMPBerry},
\begin{equation}
C_\mu = \frac{1}{2\pi} \int_{-\pi}^\pi dq \oint d\tau \left(\partial_q A_{\mu\tau} - \partial_\tau A_{\mu q}\right),\label{Ch}
\end{equation}
with the Berry connection $A_{\mu\tau} =i\langle \psi_\mu |\partial_\tau \psi_\mu \rangle$ on $\vec{\tau}$,
\begin{equation}
A_{\mu\tau} =\frac{1}{2} \frac{\partial}{\partial q}\ln (\omega_\mu^2-J_3^2)\; \frac{d}{d\tau}\ln \frac{J_1}{J_2},\label{At0}
\end{equation}
where we used the group velocity expression, $\partial\omega_\mu/\partial q = 2p\sin q/ (3\omega_\mu^2 -J^2)$, derived from \eref{disp}. \eref{At0} shows that contribution of $A_{\mu\tau}$ to~\eref{Ch} is zero on any contour $\vec{\tau}$ because the integral $\int_{-\pi}^{\pi}\partial_q A_{\mu\tau} dq$ vanishes due to periodicity $\omega(q+2\pi)=\omega(q)$. Reordering $\oint  \partial_\tau \left(\int_{-\pi}^{\pi} A_{\mu q} dq\right) d\tau$ in~\eref{Ch} we obtain
\begin{equation}\label{Chern}
C_\mu = -\frac{1}{2\pi}\oint \frac{\partial Z_\mu}{\partial \tau}d\tau,
\end{equation}
i.e. the Chern number for each band is the negative winding number of the Zak phase on the trajectory $\vec{\tau}$.

The first result of~\eref{Chern} is that Zak phase dislocations in~\fref{Fig3} define Chern numbers $C_\mu=-l_\mu =\{-1,2,-1\}$ on any contour around dislocation, thus confirming the numerical results of Ref.~\cite{ChaoYuri}. Topological nature of Chern numbers is explicit here as $C_\mu$ do not depend on a particular trajectory as long as it encircles degeneracy axis $J_0$ once in the counter-clockwise direction.

\begin{figure}[t]
\includegraphics[width=1.0\columnwidth]{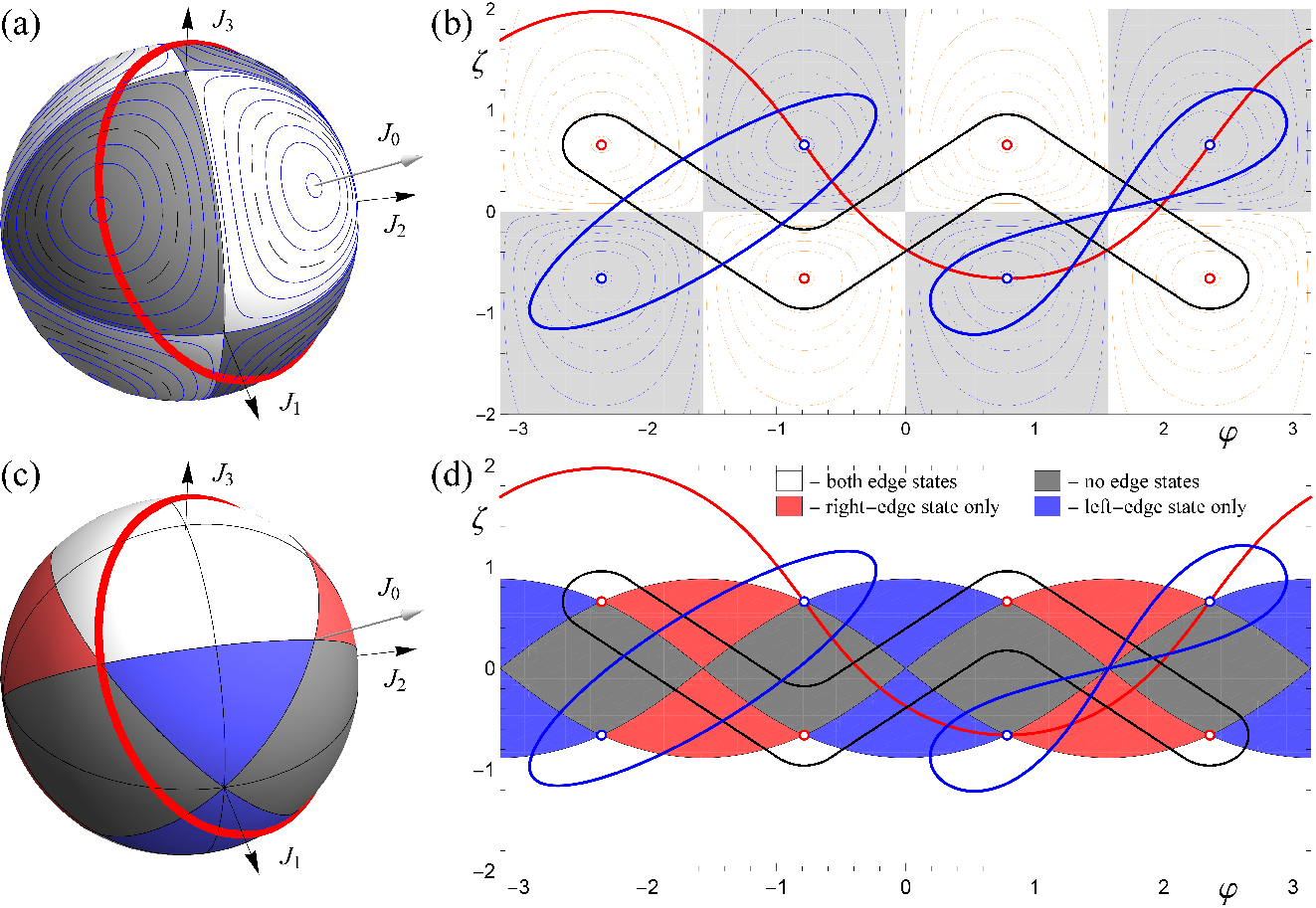}%
\caption{\label{Fig4} Full parameter space on the sphere $J$ ({\bf a,c}) and corresponding Mercator projection in ({\bf b,d}) with $\zeta = \tanh^{-1} (J_3/J)$ and spherical azimuth $\varphi =\tan^{-1}(J_2/J_1)$. In ({\bf a,b}) the white domains of ``positive'' dislocations $l_\mu=\{1,-2,1\}$, as in~\fref{Fig3} with $p>0$, and the gray domains with ``negative'' dislocations with $p<0$ and opposite winding numbers, $l_\mu=\{-1,2,-1\}$. Positive and negative dislocations are marked in ({\bf b,d}) by open red and blue circles, respectively. In ({\bf c,d}) the shading of different domains indicates the number of edge states, see the legend in panel ({\bf d}) and~\fref{Fig1}a. The solid red line in all panels, passing through 3 negative dislocations, is the AAH critical trajectory~\eref{J} with $\lambda=4$~\cite{ChaoYuri}. Other trajectories indicated in ({\bf b,d}) with black and blue lines are discussed in the text and~\fref{Fig5}.}
\end{figure}

The next open question is to understand the topological transition in AAH model with
\begin{equation}\label{J}
J_m\left(\phi\right)=\tilde{J}\left \{1-\lambda \cos\left(2\pi m/3-\phi\right)\right \},
\end{equation}
predicted to occur for modulation parameter $\lambda>4$~\cite{ChaoYuri}. The factor $\tilde{J}=J\sqrt{2/3(2+\lambda^2)}$ scales hopping parameters to the sphere $J$. For small modulation, $\lambda \ll 1$, \eref{J} describes a circle around $J_0$ axis, similar to $p=const$ loops close to $J_0$ in~\fref{Fig1}. At $\lambda=1$ the circle touches the boundary $J_m=0$ of the spherical triangle in~\fref{Fig1}a, which corresponds to the anti-continuum limit $p=0$ with collapsing bands, $\omega_\mu=J\{-1,0,1\}$, see~\aref{Ap22}.

Although hopping parameters for cold atoms~\cite{pumpSSHcold16} and coupled waveguides~\cite{diagExperPRL12} take positive values, other photonic systems offer the possibility to control hopping phases~\cite{negativeNP12,AM,negativeRingNP13}. Therefore, we extend our analysis to the complete parameter space including negative hopping parameters and distinguish ``positive'' and ``negative'' Zak phase dislocations in~\fref{Fig4}a,b. The critical parameter of topological transition $\lambda = 4$ in~\eref{J} corresponds to a circle in~\fref{Fig4} (red line) passing trough three negative dislocations. Increasing $\lambda>4$ includes four dislocations within the circle, one positive and three negative, corresponding Zak winding numbers change from $l_\mu=\{1,-2,1\}$ to $l_\mu=\{-2,4,-2\}$, and we recover the numerical result of Ref.~\cite{ChaoYuri}, $C_\mu=\{2,-4,2\}$.

\begin{figure}[t]
\begin{adjustwidth}{-\extralength}{0cm}
\includegraphics[width=1.33\textwidth]{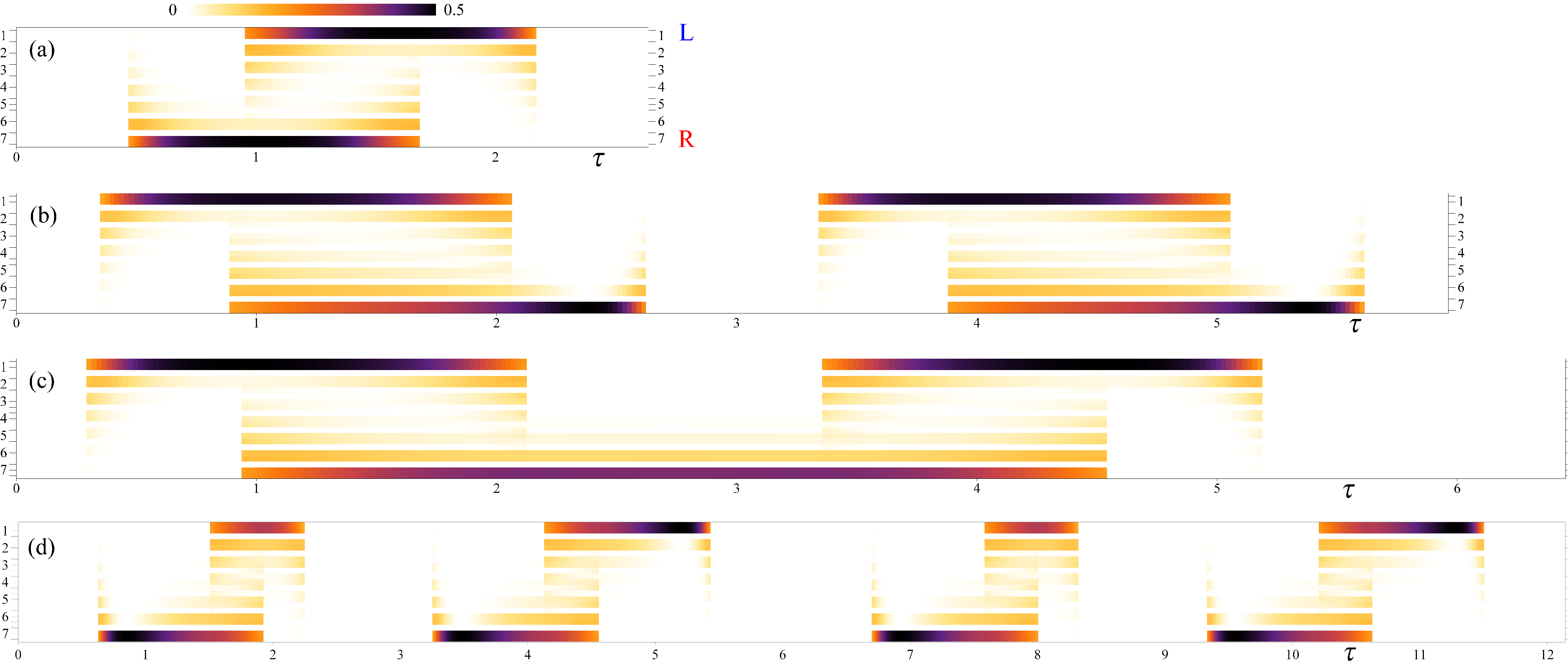}%
\caption{\label{Fig5} Total intensity of two edge-states in the bottom gap, $|\psi_R|^2+|\psi_L|^2$, along different loops, for an open chain of $N=7$ unit cells. ({\bf a}) The loop $p=0.1$ in~\fref{Fig1} (dashed contour) and~\fref{Fig2}. The oval ({\bf b}) and figure-eight ({\bf c}) trajectories shown as blue curves in~\fref{Fig4}b,d. ({\bf d}) The zig-zag shaped contour shown with a black line in~\fref{Fig4}b,d. Note that the length scale in ({\bf a-c}) is the same, which allows better comparison of the edge states and emphasises the different lengths of the loops in parameter space, important for experimental realization. The contour in ({\bf d}) is much longer and this panel length is scaled down.}
\end{adjustwidth}
\end{figure}

Mapping of Zak phase dislocations in~\fref{Fig4} offers a general way to design pumping protocols with desired Chern numbers by choosing appropriate trajectory in parameter space. For example, the same Chern numbers as above, $C_\mu=\{2,-4,2\}$, we obtain for an elliptic loop encircling two negative dislocations in~\fref{Fig4}b. The maximal Chern numbers with a single simply-connected loop are $C_\mu=\{-4,8,-4\}$ for the zig-zag shaped contour drawn with a black line in~\fref{Fig4}b, cf. Refs.~\cite{largeChern, largeChern_2,Chern4}. Furthermore, the adiabatic trajectory is not limited by simple loops, e.g. the self-intersecting figure-eight cycle in~\fref{Fig4}b has a well defined winding and Chern numbers $\{0,0,0\}$. The question we address below is how such a variety of Chern numbers relates to the adiabatic pump dynamics; namely, what is the bulk-boundary correspondence in our system.

Thouless pumping in fermionic systems is characterised by the total Chern number of the filled bands below the gap with Fermi level~\cite{Thouless}. In our case $C=C_1$ for the bottom gap, $C=C_1+C_2$ for the top gap, and $C=\sum C_\mu=0$ when all bands are full. Photonic lattices realize filled bands with excitation of a single lattice site~\cite{ChaoYuri} while they also allow precise excitation of selected modes and adiabatic pumping with edge-states~\cite{diagExperPRL12}. Therefore, in the following we discuss the bottom gap states and $C=C_1$.

\fref{Fig4}c,d maps the domains with different number of edge-states with the same coloring as in~\fref{Fig1}a, it allows to predict the bifurcation of each gap edge-state, e.g. in the gray parameter areas the gaps are empty. In~\fref{Fig5} we plot the total intensity of edge-states in the bottom gap, $|\psi_L|^2+|\psi_R|^2$, along different trajectories. The states on two edges overlap only in the $b$ sublattice because the left-edge state has sublattice $c$ empty, $\psi_L=\{a_L,b_L,0\}^T$, while the right-edge state has sublattice $a$ empty, $\psi_R=\{0,b_R,c_R\}^T$. \fref{Fig5}a corresponds to the dashed loop $p=0.1$ in~\fref{Fig1} with a sequence of ``right-both-left'' edge states along $\phi$, cf.~\fref{Fig2}b. This sequence corresponds to the Thouless pumping of a single charge from the right to the left edge~\cite{diagExperPRL12}, i.e. negative current with the total Chern number $C=C_1=-1$. We will count the pair of edge states in the sequence RL (right-left) as a negative ``Thouless pair'', in contrast to the positive pair LR (left-right). Examples below establish the correspondence: the Chern number equals the total number of (signed) edge-state pairs.

Indeed, the elliptic loop in~\fref{Fig4}b corresponds to the sequence of two positive edge-state pairs LR-LR in~\fref{Fig5}b, thus $C=2$ (similar to the circle~\eref{J} with $\lambda>4$). The figure-eight loop in~\fref{Fig4}b demonstrates the sequence LR-RL in~\fref{Fig5}c, the two opposite currents cancel each other within one cycle, thus $C=0$. The zig-zag loop with $C=-4$ has four consecutive negative pairs RL-RL-RL-RL, see~\fref{Fig5}d, and it realizes the maximal Chern numbers in our system, $C_\mu=\{-4,8,-4\}$.

A final note on the design of adiabatic pumps is that the full three-dimensional parameter space can be utilised, e.g. the self-crossing of the figure-eight trajectory in~\fref{Fig4}b and~\fref{Fig5}c can be avoided if the coupling strength $J$ varies along the cycle. In other words, the trajectories are not limited to the sphere, considered above, and they can form three-dimensional loops winding around degeneracy axes. The Chern numbers are exactly the same and the edge-states follow the same bifurcation sequences as their counterparts along the loops projected on a sphere. This degree of freedom might be useful in experimental designs.

\section{Conclusions}\label{S4}

In conclusions, we consider trimer lattices and show that the gap closing degeneracies, embedded in three-dimensional parameter space, generate screw-type dislocations in geometric phase~\cite{BerryDislocations}. The Dirac-type spectral degeneracies allow for classification of Floquet–Bloch systems~\cite{SingulNJP15}, here we provide the route to design these structures based on geometric phase dislocations. This procedure allows to construct flat bands in synthetic dimensions, as well as engineering the Chern numbers and edge-states in adiabatic pumps. It remains to be explored how our findings could be applied in closely related and rapidly developing topological systems, such as two-dimensional photonic crystals~\cite{OL24_multiple,PRL25_large,PRL25}, non-hermitian lattices~\cite{nonhermitian,CPL24,PRB25_NonH,NC25_NonH}, plasmonics~\cite{SSH3plasmon}, metamaterials~\cite{meta}, and acoustics~\cite{acousticSSH3EdgeStates,acousticExper}. Perhaps more generally, spectral singularities~\cite{mapping} and topological invariants~\cite{reflect,reflect1} map to real space phase defects, we expect similar phenomena with Zak phase dislocations. It will be also of interest to study the role which geometric phase dislocations may play in topological transitions induced by the on-site potentials~\cite{ChaoYuri} and nonlinear interactions~\cite{SymmNatCom19,NonlScience,nonlinear}.

\vspace{6pt}



\authorcontributions{Conceptualization, S.E.K and A.S.D; investigation, T.U., A.U., S.E.K and A.S.D; writing---original draft preparation, T.U. and A.S.D.; writing---review and editing, T.U. and A.S.D.; supervision, A.S.D.; project administration, A.S.D.; funding acquisition, A.S.D. All authors have read and agreed to the published version of the manuscript.}

\funding{This research was funded by Nazarbayev University grant FDCRGP 040225FD4747.}

\dataavailability{The original contributions presented in this study are included in the article. Further inquiries can be directed to the corresponding author.}

\acknowledgments{We thank Eugene Demler for stimulating discussions.}

\conflictsofinterest{The authors declare no conflicts of interest.}

\abbreviations{Abbreviations}{
The following abbreviations are used in this manuscript:
\\
\noindent
\begin{tabular}{@{}ll}
SSH & Su-Schrieffer-Heeger model\\
AAH & Aubry-André-Harper model\\
\end{tabular}
}

\appendixtitles{yes} 
\appendixstart
\appendix
\section[\appendixname~\thesection]{Parameter domain}\label{Ap1}
As we discuss in the main text and~\fref{Fig1}, it is convenient to separate the two parameters on the sphere into the dispersion parameter $p$, independent from the cyclic parameter $\phi$. We introduce new parameter space $\left\{j_1,j_2,j_3\right\}$ by rotating $\left\{J_1,J_2,J_3\right\}$ till the $j_3$ axis is parallel to $\vec{J_0}$. We also identify $j_1$ axis ($j_2=0$) with the inversion-symmetric lattice  $J_1=J_2$, as seen in~\fref{Fig1}a. Corresponding transformation
\begin{equation}
\label{A1}
\left\{\begin{matrix}J_1\\J_2\\J_3\\\end{matrix}\right\}=\frac{J}{\sqrt6}\left\{\begin{matrix}j_1 -\sqrt3j_2+\sqrt2j_3\\j_1+\sqrt3j_2+\sqrt2j_3\\{-2j}_1+\sqrt2j_3\\\end{matrix}\right\}
\end{equation}
preserves the norm $j_1^2+j_2^2+j_3^2\equiv\ j^2=J^2$. Next, we introduce cylindrical coordinates on the unit sphere, $\left\{j_1,j_2,j_3\right\}=\left\{r\cos\phi,r\sin\phi,\sqrt{1-r^2}\right\}$, so that each curve $J_1J_2J_3=p$ is parameterized with $\phi=0\ldots2\pi$ on the trajectory $r=\rho(\phi,p)$ found numerically as the solution of
\begin{equation}
\label{A3}
6\sqrt3\ p=\left(2-5\rho^2\right)\sqrt{1-\rho^2}-\sqrt2\rho^3\cos{3\phi}.
\end{equation}
These trajectories are shown with contour lines in~\fref{Fig1} and~\fref{Fig4}a,b for different values of $p$, and one particular (dashed) trajectory with $p=0.1$ is also shown in~\fref{Fig3}; the details of this family of states are presented in~\fref{Fig2} and~\fref{Fig5}a.

\section[\appendixname~\thesection]{Calculation of the Zak phase}
\subsection[\appendixname~\thesubsection]{Zak phase vanishes in the inversion-symmetric lattice $J_1=J_2$}\label{Ap21}

Although directly evident from~\eref{Aq} that Berry connection vanishes for $J_1^2=J_2^2$, we also notice that the eigenvector $\psi=\left\{a,b,c\right\}^T$ in~\eref{psi} gains additional symmetry, $c=\pm a^\ast$ for $J_2=\pm J_1$. We can generalize and consider a vector $\psi=\left\{a,b,\pm a^\ast\right\}^T$ with real  $b=b^\ast=\sqrt{1-2\left|a\right|^2\ }$. It is straightforward to derive $\langle\psi|\partial_q \psi\rangle\equiv0$ for such an eigenstate.

Furthermore, the eigenstate amplitude $\Psi_\mu$ in~\eref{psi} and~\eref{Aq} diverges in this case, which may complicate the direct numerical integration of~\eref{Zak} and~\eref{Ch}. The dispersion relation~\eref{disp} for $q=0,\pi$, plotted in~\fref{Fig4}b,c, gives the following results:

\begin{tabular}{@{}llcl}
For  &$J_3>J_1=J_2$, &$\phi=\pi$, &$\begin{cases} \omega_3(0)=J_3\\ \omega_1(\pi)=-J_3 \end{cases}$ \\[5mm]
For  &$J_3<J_1=J_2$, &$\phi=0$,   &$\begin{cases} \omega_2(0)=J_3\\ \omega_2(\pi)=-J_3 \end{cases}$\\[5mm]
\end{tabular}

\noindent
so that the factor $1/(\omega_\mu^2-J_3^2)$ diverges for respective bands. Similarly, the factor $1/(3\omega_\mu^2-J^2)$ diverges at the degeneracy $J_0$, i.e. $J_{1,2,3}=J/\sqrt{3}$, where a pair of bands touch and form Dirac cones in~\fref{Fig1}b,c, namely $\omega_2(0)=\omega_3(0)=J/\sqrt{3}$ and $\omega_1(\pi)=\omega_2(\pi)=-J/\sqrt{3}$. These poles are not essential, they appear as a consequence of band folding.

\subsection[\appendixname~\thesubsection]{Band collapse and Zak phase in the anti-continuum limit}\label{Ap22}
Along the boundary of the spherical triangle in~\fref{Fig1}a one of the hopping parameters vanishes. This corresponds to the fragmented (anti-continuum) lattice and leads to the band collapse. Indeed, because the parameter $p=0$, the dispersion relation~\eref{disp} gives the triplet $\omega_{1,2,3}=J\{-1,0,1\}$, independent of the location at the boundary. We were able to derive Zak phase in analytical form in this limit using expressions for eigenvectors~\eref{psi} and Berry connection~\eref{Aq}. We obtain the following results, referring to the edges of the spherical triangle in~\fref{Fig1}a:

\begin{tabular}{@{}llcl}
Right edge:   &$J_1=0$, &$\pi/3<\phi<\pi$,    &$\begin{cases}Z_{1,3}=\pi J_3^2/J^2 \\Z_2=2\pi J_2^2/J^2\end{cases}$ \\[5mm]
Left edge:    &$J_2=0$, &$-\pi<\phi<-\pi/3$,  &$\begin{cases}Z_{1,3}=-\pi J_3^2/J^2 \\Z_2=-2\pi J_1^2/J^2\end{cases}$\\[5mm]
Bottom edge:  &$J_3=0$, &$-\pi/3<\phi<\pi/3$, &\quad $Z_{1,2,3}=0$\\[5mm]
\end{tabular}

\noindent
These results are fully consistent with numerically obtained in~\fref{Fig3}, namely the winding numbers of Zak phase dislocations $l_{1,2,3}=\left\{1,-2,1 \right\}$. It is noteworthy that such a simple analysis allows to identify the presence of a dislocation and its winding number. In particular, it offers a quick proof that the winding numbers change sign when one of the hopping parameters above changes sign, as shown in~\fref{Fig4}.


\begin{adjustwidth}{-\extralength}{0cm}

\reftitle{References}

\PublishersNote{}
\end{adjustwidth}


\begin{thebibliography}{99}

\bibitem{BerryPhase}
Berry, M.V. Quantal Phase Factors Accompanying Adiabatic Changes \textit{Proc. R. Soc. A} \textbf{1984}, \textit{392}, 45-57, \\
https://doi.org/10.1098/rspa.1984.0023.

\bibitem{RMPBerry}
Xiao, D.; Chang, M.C.; Niu, Q. Berry phase effects on electronic properties \textit{Rev. Mod. Phys.} \textbf{2010}, \textit{82}, 1959,
\\
https://doi.org/10.1103/RevModPhys.82.1959.

\bibitem{RMPHasanKane}
Hasan, M. Z.; Kane, C. L. \textit{Colloquium}: Topological insulators \textit{Rev. Mod. Phys.} \textbf{2010}, \textit{82}, 3045,

https://doi.org/10.1103/RevModPhys.82.3045.

\bibitem{RMPtopins}
Qi, X.-L.; Zhang, S.-C. Topological insulators and superconductors \textit{Rev. Mod. Phys.} \textbf{2011}, \textit{83}, 1057, \\
https://doi.org/10.1103/RevModPhys.83.1057.

\bibitem{TP14}
Lu, L.; Joannopoulos, J. D.; Soljačić, M. Topological photonics \textit{Nat. Phot.} \textbf{2014}, \textit{8}, 821-829,
https://doi.org/10.1038/nphoton.2014.248.

\bibitem{TP19}
Ozawa, T.; Price, H. M.; Amo, A.; Goldman, N.; Hafezi, M.; Lu, L.; Rechtsman, M. C.; Schuster, D.; Simon, J.; Zilberberg, O.; et al. Topological photonics \textit{Rev. Mod. Phys.} \textbf{2019}, \textit{91}, 1, https://doi.org/10.1103/RevModPhys.91.015006.

\bibitem{TP20}
Segev, M.; Bandres, M. A. Topological photonics: Where do we go from here? \textit{Nanophotonics} \textbf{2021}, \textit{10}, 425-434, https://doi.org/10.1515/nanoph-2020-0441.

\bibitem{Zak}
Zak, J. Berry's Phase for Energy Bands in Solids \textit{Phys. Rev. Lett.} \textbf{1989}, \textit{62}, 2747-2750, https://doi.org/10.1103/PhysRevLett.62.2747.


\bibitem{ZakBloch}
Atala, M.; Aidelsburger, M.; Barreiro, J. T.; Abanin, D.; Kitagawa, T.; Demler,  E.; Bloch, I. Direct measurement of the Zak phase in topological Bloch bands \textit{Nat. Phys.} \textbf{2013}, \textit{9}, 795–800, https://doi.org/10.1038/nphys2790.


\bibitem{SSH}
Su, W. P.; Schrieffer, J. R.; Heeger, A. J. Solitons in Polyacetylene \textit{Phys. Rev. Lett.} \textbf{1979}, \textit{42}, 1698,

https://doi.org/10.1103/PhysRevLett.42.1698.


\bibitem{RM}
Rice,  M. J.; Mele, E. J. Elementary excitations of a linearly conjugated diatomic polymer \textit{Phys. Rev. Lett.} \textbf{1982}, \textit{49}, 1455, https://doi.org/10.1103/PhysRevLett.49.1455.

\bibitem{ZakSSH}
Delplace, P.; Ullmo, D.; Montambaux, G. Zak phase and the existence of edge states in graphene \textit{Phys. Rev. B} \textbf{2011}, \textit{84}, 195452, https://doi.org/10.1103/PhysRevB.84.195452.

\bibitem{diagonal12}
Lang, L.-J.; Cai, X.; Chen, S. Edge States and Topological Phases in One-Dimensional Optical Superlattices \textit{Phys. Rev. Lett.} \textbf{2012}, \textit{108}, 220401, https://doi.org/10.1103/PhysRevLett.108.220401.

\bibitem{ChaoYuri}
Ke, Y.; Qin, X.; Mei, F.; Zhong, H.; Kivshar, Yu. S.; Lee, C. Topological phase transitions and Thouless pumping of light in photonic waveguide arrays \textit{Laser Phot. Rev.} \textbf{2016}, \textit{10}, 995, https://doi.org/10.1002/lpor.201600119.


\bibitem{trimerized17}
Jin, L. Topological phases and edge states in a non-Hermitian trimerized optical lattice \textit{Phys. Rev. A} \textbf{2017}, \textit{96}, 032103, https://doi.org/10.1103/PhysRevA.96.032103.


\bibitem{ScRep17_trimer}
Liu, X.; Agarwal, G. S. The New Phases due to Symmetry Protected Piecewise Berry Phases; Enhanced Pumping and  Nonreciprocity in Trimer Lattices \textit{Sci. Rep.} \textbf{2017}, \textit{7}, 45015, https://doi.org/10.1038/srep45015.

\bibitem{Brazil}
Martinez Alvarez, V. M.; Coutinho-Filho, M. D.; Edge states in trimer lattices \textit{Phys. Rev. A} \textbf{2019}, \textit{99}, 013833, https://doi.org/10.1103/PhysRevA.99.013833.

\bibitem{BerryDislocations}
Nye, J. F.; Berry, M. V. Dislocations in wave trains \textit{Proc. R. Soc. A} \textbf{1974}, \textit{336}, 165, https://doi.org/10.1098/rspa.1974.0012.

\bibitem{Soskin}
Soskin, M. S.; Vasnetsov, M. V. Chapter 4 - Singular optics, In \textit{Progress in Optics}, 1st ed.; Wolf, E.; Elsevier: Amsterdam, Netherlands, 2001; Volume 42, pp. 219-276, ISBN: 0-444-50908-9.

\bibitem{DesyatnikovKivsharTorner}
Desyatnikov, A. S.; Kivshar, Yu. S.; Torner, L. Optical Vortices and Vortex Solitons, In \textit{Progress in Optics}, 1st ed.; Wolf, E.; Elsevier: Amsterdam, Netherlands, 2005; Volume 47, pp. 291-391, ISBN: 0444515984.

\bibitem{DennisOHolleran}
Dennis, M. R.; O'Holleran, K.; Padgett, M. J. Singular Optics: Optical Vortices and Polarization Singularities, In \textit{Progress in Optics}, 1st ed.; Wolf, E.; Elsevier: Amsterdam, Netherlands, 2009; Volume 53, pp. 293-363, ISBN: 978-0-444-53360-9.

\bibitem{OAM}
Allen, L.; Barnett, S. M.; Padgett, M. J. \textit{Optical Angular Momentum}, 1st ed.; Institute of Physics Publishing: Bristol, UK, 2003; ISBN: 9780750309011.

\bibitem{Thouless}
Thouless, D. J.; Quantization of particle transport \textit{Phys. Rev. B} \textbf{1983}, \textit{27}, 6083, https://doi.org/10.1103/PhysRevB.27.6083.

\bibitem{pumpSSHcold16}
Lohse, M.; Schweizer, C.; Zilberberg, O.; Aidelsburger, M.; Bloch, I. A Thouless quantum pump with ultracold bosonic atoms in an optical superlattice \textit{Nat. Phys.} \textbf{2016}, \textit{12}, 350-354, https://doi.org/10.1038/nphys3584.

\bibitem{pumpSSHcold16_2}
Nakajima, S.; Tomita, T.; Taie, S.; Ichinose, T.; Ozawa, H.; Wang, L.; Troyer, M.; Takahashi, Y.; Topological Thouless pumping of ultracold fermions \textit{Nat. Phys.} \textbf{2016}, \textit{12}, 296–300, https://doi.org/10.1038/nphys3622.


\bibitem{diagExperPRL12}
Kraus, Y. E.; Lahini, Y.; Ringel, Z.; Verbin, M.; Zilberberg, O. Topological States and Adiabatic Pumping in Quasicrystals \textit{Phys. Rev. Lett.} \textbf{2012}, \textit{109}, 106402, https://doi.org/10.1103/PhysRevLett.109.106402.


\bibitem{Longhi}
Longhi, S. Quantum-optical analogies using photonic structures \textit{Laser \& Photon. Rev.} \textbf{2009}, \textit{3}, 243-261,

https://doi.org/10.1002/lpor.200810055.

\bibitem{GaranovichLonghiKivshar}
Garanovich, I. L.; Longhi, S.; Sukhorukov, A. A.; Kivshar, Yu. S. Light propagation and localization in modulated photonic lattices and waveguides \textit{Phys. Rep.} \textbf{2012}, \textit{518}, 1-79, https://doi.org/10.1016/j.physrep.2012.03.005.


\bibitem{NATMoti}
Rechtsman, M. C.; Zeuner, J. M.; Plotnik, Y.; Lumer, Y.; Podolsky, D.; Dreisow, F.; Nolte, S.; Segev, M.; Szameit, A. Photonic Floquet topological insulators \textit{Nature} \textbf{2013}, \textit{496}, 196–200, https://doi.org/10.1038/nature12066.

\bibitem{nonlinearTP}
Smirnova, D.; Leykam, D.; Chong, Y.; Kivshar, Yu. S. Nonlinear topological photonics \textit{Appl. Phys. Rev.} \textbf{2020}, \textit{7}, 021306, https://doi.org/10.1063/1.5142397.


\bibitem{activeNano}
Ota, Y.; Takata, K.; Ozawa, T.; Amo, A.; Jia, Z.; Kante, B.; Notomi, M.; Arakawa,  Y.; Iwamoto, S. Active topological photonics \textit{Nanophotonics} \textbf{2020}, \textit{9}, 547, https://doi.org/10.1515/nanoph-2019-0376.

\bibitem{Nat18a}
Lohse, M.; Schweizer, C.; Price, H.; Zilberberg, O.; Bloch, I. Exploring 4D quantum Hall physics with a 2D topological charge pump \textit{Nature} \textbf{2018}, \textit{553}, 55 https://doi.org/10.1038/nature25000

\bibitem{Nat18b}
Zilberberg, O.; Huang, S., Guglielmon, J.; Wang, M.; Chen, K.P.; Kraus, Y.E.; Rechtsman, M.C. Photonic topological boundary pumping as a probe of 4D quantum Hall physics \textit{Nature} \textbf{2018}, \textit{553}, 59 https://doi.org/10.1038/nature25011

\bibitem{synthetic}
Yuan, L.; Lin, Q.; Xiao, M.; Fan, S. Synthetic dimension in photonics \textit{Optica} \textbf{2018}, \textit{5}, 1396-1405,

https://doi.org/10.1364/OPTICA.5.001396.

\bibitem{NATMotiSynth}
Lustig, E.; Weimann, S.; Plotnik, Y.; Lumer, Y.; Bandres, M. A.; Szameit, A.; Segev, M. Photonic topological insulator in synthetic dimensions \textit{Nature} \textbf{2019}, \textit{567}, 356, https://doi.org/10.1038/s41586-019-0943-7.

\bibitem{ScienceFan20}
Dutt, A.; Lin, Q.; Yuan, L.; Minkov, M.; Xiao, M.; Fan, S. A single photonic cavity with two independent physical synthetic dimensions \textit{Science} \textbf{2020}, \textit{367}, 59, https://www.science.org/doi/10.1126/science.aaz3071.


\bibitem{Andreys}
Maczewsky, L. J.; Wang, K.; Dovgiy, A. A.; Miroshnichenko, A. E.; Moroz, A.; Ehrhardt, M.; Heinrich, M.; Christodoulides, D. N.; Szameit, A.; Sukhorukov, A. A. Synthesizing multi-dimensional excitation dynamics and localization transition in one-dimensional lattices  \textit{Nat. Photon.} \textbf{2020}, \textit{14} , 76–81, https://doi.org/10.1038/s41566-019-0562-8.

\bibitem{AAH}
Aubry, S.; Andr\'{e}, G. Analyticity breaking and Anderson localization in incommensurate lattices \textit{Ann. Isr. Phys. Soc.} \textbf{1980}, \textit{3}, 18.

\bibitem{Harper}
Harper, P. G. Single Band Motion of Conduction Electrons in a Uniform Magnetic Field \textit{Proc. Phys. Soc. A} \textbf{1955}, \textit{68}, 874,
https://doi.org/10.1088/0370-1298/68/10/304.

\bibitem{Alex} Szameit, A.; Nolte, S. Discrete Optics in Femtosecond-Laser-Written Photonic Structures \textit{J. Phys. B: At. Mol. Opt. Phys.} \textbf{2010}, \textit{43}, 163001, https://doi:10.1088/0953-4075/43/16/163001.

\bibitem{LeykamDirac_2}
Leykam, D.; Desyatnikov, A. S. Conical intersections for light and matter waves \textit{Adv. Phys. X} \textbf{2016} \textit{1}, 101, https://doi.org/10.1080/23746149.2016.1144482.

\bibitem{LeykamDirac}
Diebel, F.; Leykam, D.; Kroesen, S.; Denz, C.; Desyatnikov, A. S. Conical Diffraction and Composite Lieb Bosons in Photonic Lattices \textit{Phys. Rev. Lett.} \textbf{2016}, \textit{116}, 183902, https://doi.org/10.1103/PhysRevLett.116.183902.

\bibitem{flach}
Leykam, D.; Andreanov, A.; Flach, S. Artificial flat band systems: from lattice models to experiments \textit{Adv. Phys. X} \textbf{2018}, \textit{3}, 1473052, https://doi.org/10.1080/23746149.2018.1473052.

\bibitem{flach_2}
Leykam, D.; Flach, S. Perspective: Photonic flatbands \textit{APL Photon.} \textbf{2018}, \textit{3}, 070901, https://doi.org/10.1063/1.5034365.


\bibitem{Chao19}
Hu, S.; Ke, Y.; Deng, Y.; Lee, C. Dispersion-suppressed topological Thouless pumping \textit{Phys. Rev. B} \textbf{2019}, \textit{100}, 064302, https://doi.org/10.1103/PhysRevB.100.064302.

\bibitem{inversionPRB19}
Marques, A. M.; Dias, R. G. One-dimensional topological insulators with noncentered inversion symmetry axis. \textit{Phys. Rev. B} \textbf{2019}, \textit{100}, 041104(R), https://doi.org/10.1103/PhysRevB.100.041104.

\bibitem{SSH3BulkEdgeCorr}
Anastasiadis, A.; Styliaris, G.; Chaunsali, R.; Theocharis, G.; Diakonos, F. K. Bulk-edge correspondence in the trimer Su-Schrieffer-Heeger model. \textit{Phys. Rev. B} \textbf{2022}, \textit{106}, 085109, https://doi.org/10.1103/PhysRevB.106.085109.

\bibitem{SSH3extended}
Verma, S.; Ghosh, T. K. Bulk-boundary correspondence in extended trimer Su-Schrieffer-Heeger model. \textit{Phys. Rev. B} \textbf{2024}, \textit{110}, 125424, https://doi.org/10.1103/PhysRevB.110.125424.

\bibitem{SSH3IRM}
Guo, Q.-H.; Zhang, Y.; Wan, X.-H.; Zheng, L.-Y. Isospectral reduction of the trimer Su-Schrieffer-Heeger lattice and its bulk-edge correspondence. \textit{Phys. Rev. Appl.} \textbf{2025}, \textit{23}, L031001, https://doi.org/10.1103/PhysRevApplied.23.L031001.

\bibitem{Symmetry24}
Zhang, R.; Chen, T. Symmetry-Related Topological Phases and Applications: From Classical to Quantum Regimes. \textit{Symmetry} \textbf{2024}, \textit{16}, 1673. https://doi.org/10.3390/sym16121673



\bibitem{Haldane04}
Haldane, F. D. M. Berry Curvature on the Fermi Surface: Anomalous Hall Effect as a Topological Fermi-Liquid Property. \textit{Phys. Rev. Appl.} \textbf{2004} \textit{93}, 206602, https://doi.org/10.1103/PhysRevLett.93.206602.



\bibitem{negativeNP12}
Fang, K.; Yu, Z.; Fan, S. Realizing effective magnetic field for photons by controlling the phase of dynamic modulation. \textit{Nat. Photon.} \textbf{2012}, \textit{6}, 782–787, https://doi.org/10.1038/nphoton.2012.236.

\bibitem{AM}
Poddubny, A.; Miroshnichenko, A.; Slobozhanyuk, A.; Kivshar, Y. Topological Majorana States in Zigzag Chains of Plasmonic Nanoparticles. \textit{ACS Photonics} \textbf{2014}, \textit{1}, 101–105, https://doi.org/10.1021/ph4000949.

\bibitem{negativeRingNP13}
Hafezi, M.; Mittal, S.; Fan, J.; Migdall, A.; Taylor, J. M. Imaging topological edge states in silicon photonics. \textit{Nat. Photon.} \textbf{2013}, \textit{7}, 1001-1005, https://doi.org/10.1038/nphoton.2013.274.

\bibitem{largeChern}
Skirlo, S. A.; Lu, L.; Soljačić, M. Multimode One-Way Waveguides of Large Chern Numbers \textit{Phys. Rev. Lett.} \textbf{2014}, \textit{113}, 113904, https://doi.org/10.1103/PhysRevLett.113.113904.

\bibitem{largeChern_2}
Skirlo, S. A.; Lu, L.; Igarashi, Y.; Yan, Q.; Joannopoulos, J.; Soljačić, M. Experimental Observation of Large Chern Numbers in Photonic Crystals \textit{Phys. Rev. Lett.} \textbf{2015}, \textit{115}, 253901, https://doi.org/10.1103/PhysRevLett.115.253901.

\bibitem{Chern4}
Schröter, N. B. M.; Stolz, S.; Manna, K.; de Juan, F.; Vergniory, M. G.; Krieger, J. A.; Pei, D.; Schmitt, T.; Dudin, P.; Kim, T. K.; et al. Observation and control of maximal Chern numbers in a chiral topological semimetal. \textit{Science} \textbf{2020}, \textit{369}, 179-183, https://doi.org/10.1126/science.aaz3480.

\bibitem{SingulNJP15}
Nathan, F.; Rudner, M. S. Topological singularities and the general classification of Floquet–Bloch systems. \textit{New J. Phys.} \textbf{2015}, \textit{17}, 125014, https://doi.org/10.1088/1367-2630/17/12/125014.

\bibitem{OL24_multiple} Liu, D. ; Peng, P.; Lu, X.; Shi, A.; Peng, Y.; Wei, Y.; Liu, L. Multiple topological states within a common bandgap of two non-trivial photonic crystals. \textit{Opt. Lett.} \textbf{2024}, \textit{49}, 2393, https://doi.org/10.1364/OL.520866


\bibitem{PRL25_large} Li, Z.; Li, S.; Yan, B.; Chan, H.-C.; Li, J.; Guan, J.; Bi, W.; Xiang, Y.; Gao, Z.; Zhang, S.; et al. Symmetry-Related Large-Area Corner Mode with a Tunable Mode Area and Stable Frequency. \textit{Phys. Rev. Lett.} \textbf{2025}, \textit{134}, 116607, https://doi:10.1103/PhysRevLett.134.116607.

\bibitem{PRL25} Yan, B.; Liao, B.; Shi, F.; Xi, X.; Cao, Y.; Xiang, K.; Meng, Y.; Yang, L.; Zhu, Z.; Chen, J.; et al. Realization of Topology-Controlled Photonic Cavities in a Valley Photonic Crystal. \textit{Phys. Rev. Lett.} \textbf{2025}, \textit{134}, 033803, https://doi:10.1103/PhysRevLett.134.033803.


\bibitem{nonhermitian}
Cao, S.-J.; Zheng, L.-N.; Cheng, L.-Y.; Wang, H.-F. Controllable entangled-state transmission in a non-Hermitian trimer Su-Schrieffer-Heeger chain. \textit{Phys. Rev. A} \textbf{2024}, \textit{110}, 062409, https://doi.org/10.1103/PhysRevA.110.062409.

\bibitem{CPL24} Chen, J.; Shi, A.; Peng, Y.; Peng, P.; Liu, J. Hybrid Skin-Topological Effect Induced by Eight-Site Cells and Arbitrary Adjustment of the Localization of Topological Edge States. \textit{Chin. Phys. Lett.} \textbf{2024}, \textit{41}, 037103, https://doi:10.1088/0256-307X/41/3/037103.


\bibitem{PRB25_NonH} Shi, A.; Bao, L.; Peng, P.; Ning, J.; Wang, Z.; Liu, J. Non-Hermitian Floquet Higher-Order Topological States in Two-Dimensional Quasicrystals. \textit{Phys. Rev. B} \textbf{2025}, \textit{111}, 094109, https://doi:10.1103/PhysRevB.111.094109.

\bibitem{NC25_NonH} Xie, X.; Ma, F.; Rui, W.B.; Dong, Z.; Du, Y.; Xie, W.; Zhao, Y.X.; Chen, H.; Gao, F.; Xue, H. Non-Hermitian Dirac Cones with Valley-Dependent Lifetimes. \textit{Nat. Commun.} \textbf{2025}, \textit{16}, 1627, https://doi:10.1038/s41467-025-56882-y.


\bibitem{SSH3plasmon}
Buendía, Á.; Sánchez-Gil, J. A.; Giannini, V. Exploiting Oriented Field Projectors to Open Topological Gaps in Plasmonic Nanoparticle Arrays.
\textit{ACS Photonics} \textbf{2023}, \textit{10}, 464-474, https://doi.org/10.1021/acsphotonics.2c01526.

\bibitem{meta}
Guo, Z.; Wu, X.; Ke, S.; Dong, L.; Deng, F.; Jiang, H.; Chen, H. Rotation controlled topological edge states in a trimer chain composed of meta-atoms. \textit{New J. Phys.} \textbf{2022}, 24, 063001, https://doi.org/10.1088/1367-2630/ac71bd.

\bibitem{acousticSSH3EdgeStates}
Ioannou Sougleridis, I.; Anastasiadis, A.; Richoux, O.; Achilleos, V.; Theocharis, G.; Pagneux, V.; Diakonos, F. K. Existence and characterization of edge states in an acoustic trimer Su-Schrieffer-Heeger model. \textit{Phys. Rev. B} \textbf{2024}, \textit{110}, 174311,  https://doi.org/10.1103/PhysRevB.110.174311.

\bibitem{acousticExper}
Guo, Q.-H.; Zhang, Y.; Wan, X.-H.; Zheng, L.-Y. Observation of chiral edge state pairs in an acoustic trimer waveguide. \textit{Appl. Phys. Lett.} \textbf{2025}, \textit{126}, 133102, https://doi.org/10.1063/5.0256334.


\bibitem{mapping}
Liu, X.; Xia, S.; Jajtić, E.; Song, D.; Li, D.; Tang, L.; Leykam, D.; Xu, J.; Buljan, H.; Chen, Z. Universal momentum-to-real-space mapping of topological singularities. \textit{Nat. Comm.} \textbf{2020}, \textit{11}, 1586, https://doi.org/10.1038/s41467-020-15374-x.

\bibitem{reflect}
Poshakinskiy, A. v.; Poddubny, A. N.; Hafezi, M. Phase spectroscopy of topological invariants in photonic crystals. \textit{Phys. Rev. A} \textbf{2015}, \textit{91}, 043830, https://doi.org/10.1103/PhysRevA.91.043830.

\bibitem{reflect1}
Li, Q.; Jiang, X. Singularity induced topological transition of different dimensions in one synthetic photonic system. \textit{Opt. Comm.} \textbf{2019}, \textit{440}, 32-40, https://doi.org/10.1016/j.optcom.2019.02.015.

\bibitem{SymmNatCom19}
González-Cuadra, D.; Bermudez, A.; Grzybowski, P. R.; Lewenstein, M.; Dauphin, A. Intertwined topological phases induced by emergent symmetry protection. \textit{Nat. Comm.} \textbf{2019}, textit{10}, 2694, https://doi.org/10.1038/s41467-019-10796-8.

\bibitem{NonlScience}
Maczewsky, L. J.; Heinrich, M.; Kremer, M.; Ivanov, S. K.; Ehrhardt, M.; Martinez, F.; Kartashov, Y. V.; Konotop, V. V.; Torner, L.; Bauer, D.; Szameit, A. Nonlinearity-induced photonic topological insulator. \textit{Science} \textbf{2020}, \textit{370}, 701-704, https://doi.org/10.1126/science.abd2033.

\bibitem{nonlinear}
Kartashov, Y. V.; Arkhipova, A. A.; Zhuravitskii, S. A.; Skryabin, N. N.; Dyakonov, I. V.; Kalinkin, A. A.; Kulik, S. P.; Kompanets, V. O.; Chekalin, S. V.; Torner, L.; Zadkov, V. N. Observation of Edge Solitons in Topological Trimer Arrays. \textit{Phys. Rev. Lett.} \textbf{2022}, \textit{128}, 093901, https://doi.org/10.1103/PhysRevLett.128.093901.







\end{thebibliography}
\end{document}